\documentclass[preprint,aps]{revtex4}
\usepackage{hyperref}
\usepackage{color}
\usepackage{graphicx}
\usepackage{bm}
\usepackage{amsmath}
\usepackage{hyperref}
\usepackage{enumerate}
\usepackage{upgreek}
\usepackage[subnum]{cases}

\newcommand{\erf}{ \text{erf} }

\newcommand{\pa}{ \partial }
\newcommand{\hb}{ \hbar }
\newcommand{\si}{ \sigma }

\newcommand{\ga}{ \gamma }
\newcommand{\la}{ \langle }
\newcommand{\ra}{ \rangle }

\newcommand{\al}{ \alpha }
\newcommand{\re}{ \text{Re} }
\newcommand{\sip}{ \text{sp} }

\newcommand{\mup}{ \text{m} }
\newcommand{\MB}{ \text{MB} }

\begin{document}
	
\title{On some unexplored decoherence aspects in the Caldeira-Leggett formalism: arrival time distributions, identical particles and diffraction in time}

\author{S. V. Mousavi}
\email{vmousavi@qom.ac.ir}
\affiliation{Department of Physics, University of Qom, Ghadir Blvd., Qom 371614-6611, Iran}

\author{S. Miret-Art\'es}
\email{s.miret@iff.csic.es}
\affiliation{Instituto de F\'isica Fundamental, Consejo Superior de Investigaciones Cient\'ificas, Serrano 123, 28006 Madrid, Spain}

\begin{abstract}

Some unexplored decoherence aspects within the Caldeira-Leggett master equation are analyzed and discussed. The decoherence process is controlled by the two environment parameters, the relaxation rate or friction and the temperature, leading to a gradual  transition from the quantum to classical regime. Arrival time distributions,  nonminimum-uncertainty-product or stretching Gaussian wave packets, identical particles and diffraction in time display interesting features during the decoherence process undergone by the time dependent interference patterns. We show that the presence of a constant force field does not affect the decoherence, {\it positive} values of the stretching parameter reduces the rate of decoherence, the symmetry of the wave function for identical particles plays no role when  open dynamics are considered; and diffraction in time and space is gradually washed out by increasing the temperature and/or relaxation rate in the zero dissipation limit within the so-called quantum shutter problem.

\end{abstract}

\maketitle

{\bf{Keywords}}: Decoherence; Caldeira-Leggett master equation; Arrival times; Nonminimum-uncertainty-product Gaussian wave packet; Identical particles; Diffraction in time


\section{Introduction}

Since the pioneering work by Zeh in 1970 \cite{Zeh},  decoherence has been the central subject of intense research investigation and activity although this term was 
not introduced until the late 1980s. It was Zurek \cite{Zurek} who clearly reached a milestone in this field at the beginning of the 1980s. 
Subsequently, Joos and Zeh \cite{Joos} coauthored  a seminal paper in this context. From then, an explosion of works along  this direction has not 
being stopped yet in several branches of physics as well as  chemistry and biology.
Decoherence or environment induced decoherence occurs because the system is entangled with its environment and one key aspect is the corresponding timescale due to the fact that this can be seen as a dynamical process along time. Once this process reaches a stationary state, we could claim that the transition from quantum to classical regime has been fully established.

A widely used approach to deal with open quantum systems is within the so-called {\it system-plus-environment} model where the total system is considered isolated. 
The total isolated system is described by a pure state wave function, whose evolution is determined by the unitary time evolution operator 
according to the Schr\"{o}dinger equation.  The reduced density matrix in the coordinate representation  describing the system 
of interest is obtained by tracing out the the degrees of freedom of the environment. In this way, a master equation is derived for the evolution 
of the reduced density which contains both frictional and thermal effects due to the environment, the so-called Caldeira-Leggett (CL) master equation \cite{CaLe-PA-1983,Caldeira-book-2014}. Time evolution of coherences is also an indication of how the decoherence process is established leading to certain timescales of the system under study, spatial interference terms are exponential suppressed at a given rate. As is known, when using initial Gaussian wave packets under the presence of up to quadratic interaction potentials, the width of the Gaussian in the off-diagonal elements quantifies the range of spatial coherence \cite{Schlosshauer} speaking about the coherence length.
This Markovian master equation has been used in many branches of physics such as quantum optics, quantum computation, mesoscopic systems, etc.

Alternative but less known approaches can also be found in the literature within the so-called Caldirola-Kanai and Scr\"odinger-Langevin frameworks \cite{MoMi} where the wave function of the open quantum system is considered instead of the corresponding density matrix. Both approaches are not following the system-plus-environment model but effective time dependent  Hamiltonians and nonlinear Schr\"odinger equations, respectively. 
Recently, the Caldirola-Kanai and CL approaches have been applied to describe interference and diffraction of identical particles in one slit problems \cite{MoMi-EPJP-2020-1}. 
The CL approach  stresses different features of the dynamics of open quantum systems with respect to those are using wave functions. 
Recently, this type of analysis has been carried out in the momentum space \cite{MoMi-Entropy-2021}.

In this work, the gradual decoherence process is analyzed for several problem encountered in open dynamics controlled by the two environment parameters, the relaxation rate or friction and the temperature. Thus, the quantum-to-classical transition is studied.
We show that the presence of a constant force field does not affect the decoherence. Minimum- and 
non-minimum-uncertainty-product or stretching Gaussian wave packets are used in order to show that the so-called stretching parameter reduces the rate of decoherence in the interference patterns. Time arrival distributions obtained within the Bohmian framework are first presented, identical particles  and the so-called
diffraction in time \cite{Mo-PR-1952} are analyzed. Interference patterns as well as the oscillations observed in diffraction which are the hallmark of the quantum shutter problem is gradually washed out by increasing the temperature in the zero dissipation limit.

This paper is organized as follows. In Section II the CL master equation in the coordinate representation is briefly introduced stressing the way of how arrival time distributions can be obtained. In Section III, the dynamics of non-minimum-uncertainty-product Gaussian wave packets is analyzed in order to show the importance of the so-called stretching parameter in the decoherence process. Once this analysis is carried out, the superposition of two Gaussian wave packets (the Schr\"odinger cat state) is considered  showing how the interference pattern is blurred by the presence of dissipation and temperature. Two identical particles dynamics is then briefly analyzed in Section IV. Once the interference process in space is discussed the next step is to see how the so-called diffraction in time effect is affected by the environment within the quantum shutter problem in Section V. Finally, in Section VI and VII, results and discussion and concluding remarks are presented, respectively.

\section{The Caldeira-Leggett master equation in the coordinate representation}
\label{sec: CL}

Within this approach, the environment is generally modelled by a bosonic bath, consisting of an infinite number of quantum oscillators in thermal
equilibrium, and affects the system of interest through a position-position coupling. 
The nature of this  environment model is actually known as a minimal one \cite{Caldeira-book-2014}. 
The corresponding Markovian master equation in the coordinate representation for one dimension, at the high temperature limit, is written for a particle of mass $m$ as \cite{CaLe-PA-1983, Caldeira-book-2014}
\begin{eqnarray} \label{eq: CL eq}
\frac{\pa \rho(x, x', t)}{\pa t} &=& \left[ - \frac{\hb}{2mi} \left( \frac{\pa^2}{\pa x^2} - \frac{\pa^2}{\pa x'^2} \right) - \ga (x-x') \left( \frac{\pa}{\pa x} - \frac{\pa}{\pa x'} \right)
+ \frac{ V(x) - V(x') }{ i\hb } \right. \nonumber \\
&-& \left.  \frac{D}{\hb^2} (x-x')^2 \right] \rho(x, x', t)  
\end{eqnarray}
where $V$ is the external interaction potential and 
\begin{eqnarray} \label{eq: D}
D &=& 2 m \ga k_B T  
\end{eqnarray}
plays the role of the diffusion coefficient; $k_B$ and $T$ being Boltzmann’s constant and the environment temperature, respectively. 
In the center of mass and relative coordinates
\begin{numcases}~
R = \frac{x+x'}{2} \label{eq: cm} \\
r = x - x'  \label{eq: rel} 
\end{numcases}
Eq. (\ref{eq: CL eq}) is expressed as
\begin{eqnarray} \label{eq: CLeq_Rr}
\frac{\pa \rho(R, r, t) }{\pa t} + \frac{\pa j}{\pa R} + \frac{ V(R/2+r) - V(R/2-r) }{ i\hb } + 2 \ga r \frac{\pa \rho(R, r, t) }{\pa r} + r^2 \frac{D}{\hb^2} \rho(R, r, t)  &=& 0
\end{eqnarray}
where we have defined the current density matrix as
\begin{eqnarray} \label{eq: cur_den_mat}
j(R, r, t) &=& - i \frac{\hb}{m} \frac{\pa}{\pa r} \rho(R, r, t) .
\end{eqnarray}
As is known, diagonal elements of the density matrix has the interpretation of the probability density. By imposing the condition $r=0$ in 
Eq. \ref{eq: CLeq_Rr}, one has the continuity equation
\begin{eqnarray} \label{eq: con_CL}
\frac{\pa P(x, t)}{\pa t} + \frac{\pa J(x, t)}{\pa x}  &=& 0 ,
\end{eqnarray}
where $P(x, t)$ and $J(x, t)$ are the diagonal elements, $r=0$, of $\rho(R,r,t)$ and $j(R,r,t)$, respectively. 
From the current density of probability is possible to extract information on time distributions at the screen position $x=X$ according to \cite{Mu-Lea-PR-2000}
\begin{eqnarray} \label{eq: ardis}
	\Pi_a(X, t) &=& \frac{ |J(X, t)| }{ \int_0^{\infty} dt' |J(X, t')| }
\end{eqnarray}
from which the mean arrival time and rms width of the arrival time distribution are obtained
\begin{eqnarray} 
\uptau_a(X) &=& \la t \ra = \int_0^{\infty} dt ~ t ~\Pi_a(X, t) \label{eq: meanar} \\
\si_a(X) &=& \sqrt{ \la t^2 \ra  - \la t \ra^2 } = \sqrt{ \int_0^{\infty} dt ~ ( t - \la t \ra)^2  ~\Pi_a(X, t) } \label{eq: sigma_ar}
\end{eqnarray}

This analysis can also be accompanied by a description in terms of trajectories.
Within  Bohmian mechanics where a complete description of a quantum system is given by its position and wavefunction, trajectories are computed by integrating the guidance equation \cite{Holland-book-1993, DuGoZa-JSP-1992}
\begin{eqnarray} \label{eq: guidance}
	\dot{x}(x, t) = \frac{J(x, t)}{P(x, t)}\bigg|_{x = x(x^{(0)}, t)}
\end{eqnarray}
$ x^{(0)} $ being the initial position of the Bohmian particle.
In the context of standard quantum mechanics, the concept of arrival time is ambiguous in contrast to  classical mechanics where  trajectories are well defined \cite{Mu-Lea-PR-2000}. 
Different approaches have been applied to deal with this time; approaches based on trajectories, quantization rules, time operators, phase-space techniques, renewal equations and operational procedures \cite{MuSaPa-SM-1998}. In the Bohmian context, the arrival time distribution at the screen position $x=X$ is given by the modulus of the probability current density, suitably normalized, \cite{Le-LN-2008}

\section{Dynamics of Gaussian wave packets}

In this section we first consider open dynamics under the presence or not of a linear potential for a non-minimum-uncertainty-product or {\it stretched} Gaussian wave packet in the CL framework 
and then a pure initial state consisting of superposition of two Gaussian wave packets i.e., a Schr\"{o}dinger cat state. This section will finish with two-identical-particle systems. 

\subsection{Non-minimum-uncertainty-product Gaussian wave packet}

Let us consider first an initial state given by the non-minimum-uncertainty-product Gaussian wave packet 
\begin{eqnarray} \label{eq: wf0}
\psi_0(x) &=& \frac{1}{(2\pi \si_0^2(1+i \eta)^2)^{1/4}} \exp \left[ - \frac{(x-x_0)^2}{4\si_0^2(1+i \eta)} + i \frac{p_0}{\hb} x \right],
\end{eqnarray}
where $x_0$, $p_0$ and $\sigma_0$ are the initial values for the center, momentum and width, respectively. The $\eta$ parameter is known as the 
stretching parameter and governs the uncertainty product $ \Delta x \Delta p $ with $ \Delta x = \si_0 \sqrt{1 + \eta^2}$ and 
$\Delta p = \hb / 2 \si_0 $  which are the uncertainties in position and momentum, respectively.
Eq. (\ref{eq: CL eq}) is solved under the presence of the linear potential $ V(x) = m g~x $ and coordinates (\ref{eq: cm}) and (\ref{eq: rel}) 
in three steps; first, by applying the technique of the partial Fourier transform with respect to the center of mass coordinate; second, by solving the 
resulting equation and finally taking the inverse Fourier transform of  this solution in order to obtain the density matrix in the position representation 
\cite{Ve-PRA-1994, VeKuGh-PA-1995}. In this way, one obtains
\begin{eqnarray}
\rho(R, r, t) &=& \frac{1}{ \sqrt{2\pi} w_t } \exp\left[ a_0(r, t) - \frac{ ( R - a_1(r, t) )^2 }{ 2 w_t^2} \right] \label{eq: denmat}  \\
j(R, r, t) &=& -i\frac{\hb}{m} \left( \frac{\pa a_0}{\pa r} + \frac{ x-a_1(0, t) }{w_t^2} \frac{\pa a_1}{\pa r} \right) \rho(R, r, t) \label{eq: curmat}
\end{eqnarray}
for the non-diagonal elements of the density matrix and the current density matrix, where we have defined
\begin{eqnarray}
a_0(r, t)& =& - \left[ \frac{ e^{-4\ga t} }{ 8 \si_0^2 } + \frac{ 1 - e^{-4\ga t} }{4\ga} \frac{D}{\hb^2} \right]r^2
+ i \left( \frac{p_0}{\hb} ~e^{-2\ga t}  - \frac{mg}{\hb} \uptau(t) \right) r
\label{eq: a0}
	\\
a_1(r, t) &=& x_t  
+ i \left[ \frac{ \hb }{ 4 m \si_0^2 } e^{-2\ga t} \uptau(t) + \frac{D}{m \hb} \uptau(t)^2 + \frac{\eta}{2} e^{-2\ga t} \right] r
\label{eq: a1} \\ 
w_t &=& \si_0 \sqrt{1 + \frac{ \hb^2 }{ 4 m^2 \si_0^4 } \uptau(t)^2
+ \frac{ 4\ga t + 4 e^{-2\ga t} - 3 - e^{-4\ga t} }{8 m^2 \ga^3 \si_0^2} ~ D
+ \eta \frac{\hb}{m\si_0^2} \uptau(t) + \eta^2 }  
\label{eq: wt}
\end{eqnarray}
with
\begin{eqnarray} \label{eq: tau}
\uptau(t) &=& \frac{1-e^{-2\ga t}}{2\ga} ,
\end{eqnarray}
and
\begin{eqnarray}\label{eq: xt}
x_t &=& x_0 + \frac{p_0}{m} \uptau(t) - g \frac{2\ga t-1+e^{-2\gamma t}}{4\gamma^2} .
\end{eqnarray}
By imposing the conditions $r=0$, the probability density (PD) and the probability current density (PCD)  are finally expressed as 
\begin{eqnarray}
P(x, t)& =& \frac{1}{ \sqrt{2\pi} w_t } \exp\left[ - \frac{ ( x - x_t )^2 }{ 2 w_t^2} \right] \label{eq: probden}  \\
J(x, t) &=& \left\{ \frac{p_0}{m} ~e^{-2\ga t}  - g \uptau(t) 
 \frac{ x - x_t }{w_t^2}
\left[ \uptau(t) \left( \frac{ \hb^2 }{ 4 m^2 \si_0^2 } e^{-2\ga t} + \frac{D}{m^2} \uptau(t)  \right) + \eta \frac{\hb}{2m} e^{-2\ga t} \right] \right \} \nonumber \\
 & \times & P(x, t) 
\label{eq: cur}
\end{eqnarray}
respectively.
Thus,  the PD has a Gaussian shape with a width $w_t$ and a center moving along the {\it classical} trajectory $x_t$ given by Eqs. (\ref{eq: wt}) and (\ref{eq: xt}), respectively. Notice that the stretching parameter also contributes to the spreading of the PD in a significant way.

In the free-friction limit, where the second term in the right hand side of Eq.(\ref{eq: CL eq}) i.e., the term proportional to the damping 
constant is neglected, Eqs. (\ref{eq: xt}) and (\ref{eq: wt}) reduce to
\begin{numcases}~
x_t \approx x_0 + \frac{p_0}{m} t - \frac{1}{2}g t^2 , \label{eq: xt_limit} \\
w_t \approx \si_0 \sqrt{1 + \frac{ \hb^2 }{ 4 m^2 \si_0^4 } t^2 + \frac {2 D }{ 3 m^2 \si_0^2 } ~ t^3 + \eta \frac{\hb}{m\si_0^2} t + \eta^2}  . \label{eq: wt_limit}
\end{numcases}
Note that this limit corresponds to short times where $ \ga t \ll 1 $ i.e., times much shorter than the relaxation time, $\ga^{-1}$ \cite{Ye-PRA-2010}.  
In this open dynamics, the temperature is involved through the diffusion coefficient.

\subsection{Superposition of two Gaussian wave packets; the Schr\"{o}dinger cat state}

Let us consider now the initial state as a superposition of two wave packets to be defined later on,
\begin{eqnarray} \label{eq: sup0}
\psi_0(x) &=& \mathcal{N} ( \psi_{0a}(x) + \psi_{0b}(x) ) 
\end{eqnarray}
$ \mathcal{N} $ being the normalization constant.
Equation (\ref{eq: sup0}) shows that the initial density matrix has the form
\begin{eqnarray} \label{eq: rho0}
\rho(x, x', 0) &=& \mathcal{N}^2 ( \rho_{aa}(x, x', 0) + \rho_{ab}(x, x', 0) + \rho_{ba}(x, x', 0) + \rho_{bb}(x, x', 0) ) 
\end{eqnarray}
where $ \rho_{ij}(x, x', 0) = \psi_{0i}(x) \psi_{0j}^*(x') $; $i$ and $j$ being $a$ or $b$. Due to the linearity of the master equation (\ref{eq: CL eq}), one obtains again the evolution of each term of 
Eq. (\ref{eq: rho0}) separately by using the method outlined above. Afterwards, these solutions are superposed to have the time 
dependent  PD  according to \cite{MoMi-EPJP-2020-2}, 
\begin{eqnarray} \label{eq: probden_sup}
	P(x, t) &=& \mathcal{N}^2 ( P_{aa}(x, t) + P_{ab}(x, t) + P_{ba}(x, t) + P_{bb}(x, t) )   .
\end{eqnarray}
By using the fact that $ P_{ba}(x, t) = P_{ab}^*(x, t) $, one can write
\begin{eqnarray} \label{eq: probden_sup1}
P(x, t) &=& \mathcal{N}^2 ( P_{aa}(x, t) + P_{bb}(x, t) + 2 |P_{ab}(x, t)| \cos \Theta(x, t) ) 
\end{eqnarray}
where $|P_{ab}(x, t)|$ is the modulus of $P_{ab}(x, t)$ and $ \Theta(x, t) $  its phase. Rewriting Eq. (\ref{eq: probden_sup1}) as the typical interference pattern expression \cite{BaPE-book-2002}
\begin{eqnarray} \label{eq: probden_sup2}
P(x, t) &=& \mathcal{N}^2 ( P_{aa}(x, t) + P_{bb}(x, t) + 2 \sqrt{ P_{aa}(x, t) P_{bb}(x, t) } ~ e^{\Gamma(t)} 
\cos \Theta(x, t) ) 
\end{eqnarray}
one has that
\begin{eqnarray} 
\Gamma(t) &=& \log \frac{|P_{ab}(x, t)|}{ \sqrt{ P_{aa}(x, t) P_{bb}(x, t) } } 
\end{eqnarray}
$\Gamma(t)$ being the so-called decoherence (negative) function. The  corresponding exponential function 
\begin{eqnarray} \label{eq: atten}
a(t) &=& e^{\Gamma (t)}
\end{eqnarray} 
is called the coherence attenuation coefficient which quantifies the reduction of the interference contrast \cite{FC-PLA-2001}.

\subsubsection{Minimum-uncertainty-product Gaussian wavepackets}

Let us assume that the initial state is a superposition of two wave packets located symmetrically around the origin, having the same width but opposite kick momenta, $\psi_{0a}(x)$ and $\psi_{0b}(x)$, respectively
\begin{eqnarray} \label{eq: wf0_sup} 
\psi_0(x) &=& \mathcal{N}_{\mup} \frac{1}{(2\pi \si_0^2)^{1/4}} \left\{ \exp \left[ - \frac{(x-x_0)^2}{4\si_0^2} + i \frac{p_0}{\hb} x \right] + \exp \left[ - \frac{(x+x_0)^2}{4\si_0^2} - i \frac{p_0}{\hb} x \right] \right\} 
\end{eqnarray}
where the normalization constant $ \mathcal{N}_{\mup} $ is given by 
\begin{eqnarray} \label{eq: nor}
\mathcal{N}_{\mup} &=& \left\{ 2 + 2 \exp \left[ - \frac{x_0^2}{2\si_0^2} - \frac{2 p_0^2 \si_0^2}{\hb^2}  \right] \right\}^{-1/2} .
\end{eqnarray}
Then, after lengthy but quite straightforward calculations, one obtains
\begin{numcases}~ 
\Gamma_{\mup}(t) = -\left( \frac{x_0^2}{2\si_0^2} + 2 \frac{p_0^2}{\hb^2} \si_0^2 \right) 
\left( 1 - \frac{\si_0^2}{\si_t^2} \left[ 1 + \frac{\hb^2}{4m^2\si_0^4} \uptau(t)^2 \right] \right ) \label{eq: deco_func}
	\\
\Theta_{\mup}(x, t) = \frac{ \beta_{\mup}(t)( x - \alpha_{\mup}(t) ) }{\si_t^2} \label{eq: phase}
\end{numcases}
for the decoherence function $\Gamma(t)$ and the phase function $\Theta(x, t)$ respectively, with 
\begin{numcases}~ 
\al_{\mup}(t) = \frac{g}{2\ga} ( \uptau(t) - t ) \label{eq: alpha}
\\
\beta_{\mup}(t) = - x_0 \frac{\hb}{2m\si_0^2} \uptau(t) - 2 \frac{p_0}{\hb} \si_0^2 \label{eq: beta}   
\end{numcases}
and the temperature is again present by means of the diffusion coefficient through $\si_t$ which is given by Eq. (\ref{eq: wt}) by imposing $\eta=0$ i.e., $ \si_t = w_t|_{\eta=0} $. The second term inside the second factor in Eq. (\ref{eq: deco_func}) is less than one and therefore  the decoherence function (\ref{eq: deco_func}) is essentially negative.
Eq. (\ref{eq: deco_func}) clearly shows that $ \Gamma_{\mup}(t) = 0 $ for $D=0$. This is a known result implying that the last term in Eq. (\ref{eq: CL eq}) is responsible for decoherence \cite{Zu-PT-1991}. 
As one expects, the decoherence function becomes zero for $\ga=0$. In the limit $\ga \rightarrow 0$, the CL equation converts to 
the usual Schr\"{o}dinger equation for closed systems where the superposition of two wave packets remains coherent forever. 
Notice that the decoherence function is independent of $g$ which means the applied constant force does not affect the decoherence 
function. In the presence of the applied constant force, the interference pattern is shifted by the value $ \al_{\mup}(t) $. 
From the behavior of $\si_t$ at long times, $ \ga t \gg 1 $, i.e., $ \si_t \sim \sqrt{t} $, one sees from Eq. (\ref{eq: deco_func}) 
that the decoherence function approaches the value $ - ( x_0^2 / 2\si_0^2 + 2 p_0^2 / (\hb^2 \si_0^2) )  $, that is, $ \exp[\Gamma(t)] $ 
approaches  $ \la \psi_{0b} | \psi_{0a} \ra $, the overlap of the initial states. For widely separated states, $ x_0 \gg \si_0 $, this 
overlap is extremely small meaning practically zero coherence at long times. Finally, the above general arguments and trends can be particularized to the free motion case ($g=0$).

\subsubsection{Non-minimum-uncertainty-product Gaussian wave packet in free space}

If the initial state is now a superposition of two stretched Gaussian wave packets with the same width, located symmetrically around the origin
with opposite kick moment, $\psi_{0a}(x)$ and $\psi_{0b}(x)$, respectively 
\begin{eqnarray} 
\psi_0(x) &=& \mathcal{N} \frac{1}{(2\pi \si_0^2(1+i \eta)^2)^{1/4}} \left\{
\exp \left[ - \frac{(x-x_0)^2}{4\si_0^2(1+i \eta)} + i \frac{p_0}{\hb} x \right]
+ \exp \left[ - \frac{(x+x_0)^2}{4\si_0^2(1+i \eta)} - i \frac{p_0}{\hb} x\right]
 \right\} 
 \nonumber \\
 \label{eq: wf0_sup_non} 
\end{eqnarray}
where the normalization constant $ \mathcal{N} $ is now given by 
\begin{eqnarray} \label{eq: nor}
\mathcal{N} &=& \left\{ 2 + 2 \exp \left[ - \frac{x_0^2}{2\si_0^2} + 2\frac{p_0 x_0}{\hb} \eta - \frac{2 p_0^2 (1+\eta^2) \si_0^2}{\hb^2}  \right] \right\}^{-1/2}  
\end{eqnarray}
one readily obtains
\begin{eqnarray} \label{eq: deco_func_st}
\Gamma(t) &=& \Gamma_0(t) + \eta \left( - \frac{2p_0 ( \hb x_0 + \eta p_0 \si_0^2 ) }{\hb^2} + \frac{f(t)}{2 \hb^2 m^2 \si_0^2 w_t^2} \right)
\end{eqnarray}
and
\begin{numcases}~
\Theta(x, t) =  \frac{ \beta(t)( x - \alpha_m(t) ) }{w_t^2} 
\\
\beta(t) = \beta_m(t) - \left( x_0 + \frac{p_0}{m} \uptau(t) \right) \eta - \frac{2p_0 \si_0^2}{\hb} \eta^2 
\end{numcases}
for the decoherence function and phase respectively with
\begin{eqnarray}
\Gamma_0(t) &=& -\left( \frac{x_0^2}{2\si_0^2} + 2 \frac{p_0^2}{\hb^2} \si_0^2 \right) 
\left\{ 1 - \frac{\si_0^2}{w_t^2} \left[ 1 + \frac{\hb^2}{4m^2\si_0^4} \uptau(t)^2 \right] \right \} 
\\
f(t) &=& \hb^3 x_0 [mx_0 + p_0 \uptau(t)] \uptau(t) + \eta \hb^2  [ m^2 x_0^2 + 4 m x_0 p_0 \uptau(t) +p_0^2 \uptau(t)^2 ] \si_0^2
\nonumber \\
&+& 4 (1+\eta^2) \hb m p_0 [mx_0 + p_0 \uptau(t)] \si_0^4 
+ 4 \eta (2+\eta^2) m^2 p_0^2  \si_0^6
\end{eqnarray}
Equation (\ref{eq: deco_func_st}) is too complicated to be analyzed. 
However, for motionless wave packets where $p_0=0$, the decoherence function takes a simpler form
\begin{eqnarray} \label{eq: deco_func_st_ml}
\Gamma(t) &=& - \frac{x_0^2}{2\si_0^2}
\left\{ 1 - \frac{\si_0^2}{w_t^2} \left[ 1 + \frac{\hb^2}{4m^2\si_0^4} \uptau(t)^2 \right] \right \}
+ \frac{x_0^2}{2w_t^2} \left( \eta^2 + \eta \frac{\hb}{m\si_0^2} \uptau(t) \right) 
\end{eqnarray}
showing that for {\it positive} values of the stretching parameter $ \eta $ the decoherence function is less negative leading to reduction of the rate of the decoherence.
From Eq. (\ref{eq: deco_func_st_ml}) one sees again that $ \Gamma(t) = 0 $ for $D=0$ implying that the last term in Eq. (\ref{eq: CL eq}) is responsible for decoherence \cite{Zu-PT-1991}. 
In the zero dissipation limit one obtains
\begin{eqnarray} \label{eq: decoh_func__}
\Gamma(t) &\approx& - \frac{ 4D x_0^2 t^3 }{ 12m^2\si_0^4(1+\eta^2)+12m\hb\eta \si_0^2 t + 3\hb^2 t^2 + 8 D \si_0^2 t^3 }
\end{eqnarray}
where in the limit $ \si_0 \ll x_0 $ reduces
\begin{eqnarray} \label{eq: decoh_func_}
\Gamma(t) &\approx& - \frac{t}{\tau_D}, \qquad \tau_D = \frac{3 \hb^2}{2m\ga k_B T d^2} 
\end{eqnarray}
$ \tau_D $ being the decoherence time which depends on the temperature, relaxation constant and  separation between the two initial wavepackets, $ d = 2 x_0 $ \cite{QuVe-JMPB-2008}. Thus, in this limit, the stretching parameter $\eta$ does not affect the decoherence time.

\section{Two identical particles}

For a system of two identical {\it spinless} particles, the state of the system must be (anti-)symmetric for identical (fermions) bosons under the exchange of particles. If the initial pure state is given by
\begin{eqnarray} \label{eq: Psi_pm}
\Psi_{\pm}(x_1, x_2, 0) &=& \mathcal{N}_{\pm} \{ \psi(x_1, 0) \phi(x_2, 0) \pm \phi(x_1, 0) \psi(x_2, 0) \}
\end{eqnarray}
$\psi$ and $\phi$ being one-particle wave functions, then the evolution under the two-particle CL equation yields \cite{MoMi-EPJP-2020-1}
\begin{eqnarray} \label{eq: denmat_2p}
\rho_{\pm}(x_1, x_2; x_1', x_2', t) &=& \mathcal{N}_{\pm}^2 \{ \rho_{11}(x_1; x_1', t) \rho_{22}(x_2; x_2', t) 
+ \rho_{22}(x_1; x_1', t) \rho_{11}(x_2; x_2', t) 
\nonumber \\
& & \qquad\pm \rho_{12}(x_1; x_1', t) \rho_{21}(x_2; x_2', t) \pm \rho_{21}(x_1; x_1', t) \rho_{12}(x_2; x_2', t)
\}
\end{eqnarray}
where
\begin{numcases}~
\rho_{11}(x, x', 0) = \psi_0(x) \psi_0^*(x')
\\
\rho_{22}(x, x', 0) = \phi_0(x) \phi_0^*(x')
\\
\rho_{12}(x, x', 0) = \psi_0(x) \phi_0^*(x')
\\
\rho_{21}(x, x', 0) = \phi_0(x) \psi_0^*(x')
\end{numcases}
Note that although $ \rho_{11}(x, x', t) $ and $ \rho_{22}(x, x', t) $ are one-particle densities, $ \rho_{12}(x, x', t) $ and $ \rho_{21}(x, x', t) $ are not. However, all these functions are solutions of one-particle CL equation (\ref{eq: CL eq})  satisfying the continuity equation (\ref{eq: con_CL}).
Joint detection probabilities are given by the diagonal elements of (\ref{eq: denmat_2p}); 
\begin{eqnarray} 
P_{\pm}(x_1, x_2, t) &=& \mathcal{N}_{\pm}^2 [ P_{11}(x_1, t) P_{22}(x_2, t) 
+ P_{22}(x_1, t) P_{11}(x_2, t) \pm 2\re\{ P_{12}(x_1, t) P_{21}(x_2, t)\} ]
 \nonumber \\
 \label{eq: P_2p}
\end{eqnarray}
where
\begin{eqnarray}
P_{ij}(x, t) &=& \rho_{ij}(x, x, t)
\end{eqnarray}
For distinguishable particles obeying Maxwell-Boltzmann statistics, the probability density is given by
\begin{eqnarray} \label{eq: P_2p_MB}
P_{\MB}(x_1, x_2, t) &=& \frac{1}{2} [ P_{11}(x_1, t) P_{22}(x_2, t) + P_{22}(x_1, t) P_{11}(x_2, t) ]
\end{eqnarray}
The last term of (\ref{eq: P_2p}) is due to particle indistinguishability. In this context, and due to the environment, this term becomes zero along time and we have decoherence in the sense of indistinguishability loss. 

%
For the single-particle density $ P_{\sip, \pm}(x, t) = \int_{-\infty}^{\infty} dx_2 \rho_{\pm}(x, x_2; x, x_2, t) $, one obtains
\begin{eqnarray} \label{eq: P_sp}
P_{\sip, \pm}(x, t) &=& \mathcal{N}_{\pm}^2 [  
P_{11}(x, t) + P_{22}(x, t) \pm 2 \re\{ P_{12}(x, t) s(t) \}] 
\end{eqnarray}
where the overlapping integral is
\begin{eqnarray}
s(t) &=& \int_{-\infty}^{\infty} dx' P_{21}(x', t)
\end{eqnarray}
In Appendix \ref{app: con-eq}, a continuity equation has been derived for the single-particle density.
Notice that due to the continuity equation (\ref{eq: con_CL}), $s(t)$ is independent of time and does not depend on  environment parameters $\ga$ and $T$; $ s(t) = \int dx' P_{21}(x', t) = \int dx' P_{21}(x', 0) = \la \phi(0) | \psi(0) \ra $.

If the system is isolated, states evolve under the Schr\"{o}dinger equation and we have 
\begin{eqnarray} \label{eq: P_sp_Sch}
P_{\sip, \pm}(x, t) &=& \mathcal{N}_{\pm}^2 [  
|\psi(x, t)|^2 + |\phi(x, t)|^2 \pm 2 \re\{ \la \phi(0) | \psi(0) \ra \psi^*(x, t) \phi(x, t) \}] 
\end{eqnarray}
%
Comparison of (\ref{eq: P_sp}) and (\ref{eq: P_sp_Sch}) reveals that in  open systems the quantity $ P_{12}(x, t) $ plays the role of $ \psi^*(x, t) \phi(x, t) $. Thus, in analogy to  Eq. (\ref{eq: probden_sup2}) we have again 
\begin{eqnarray}
|P_{12}(x, t)| &=& \sqrt{P_{11}(x, t) P_{22}(x, t)} e^{\Gamma_{12}(t)}
\end{eqnarray}
leading to
\begin{eqnarray} \label{eq: Gamma12}
\Gamma_{12}(t) &=& \log \frac{|P_{12}(x, t)|}{ \sqrt{ P_{11}(x, t) P_{22}(x, t) } } 
\end{eqnarray}
Taking now one-particle states $\phi$ and $\psi$ as minimum-uncertainty-product Gaussian wave packets i.e., as (\ref{eq: wf0}) $\eta=0$, with parameters $\bar{x}_0$, $\bar{\si}_0$, $\bar{p}_0$ and $x_0$, $\si_0$, $p_0$ respectively, one obtains
\begin{eqnarray} \label{eq: P_12}
P_{12}(x, t) &=& \sqrt{ \frac{ 2\si_0 \bar{\si}_0 }{ \si_0^2 + \bar{\si}_0^2 } } \frac{1}{2\sqrt{\pi b_2(t)}}
\exp \left[ b_0 - \frac{ (x - b_1(t))^2}{ 4 b_2(t) }  \right]
\end{eqnarray}
where
\begin{eqnarray}
b_0 &=&  - \frac{ \hb^2 ( x_0 - \bar{x}_0 )^2 + 4 ( p_0 - \bar{p}_0 )^2 \si_0^2 \bar{\si}_0^2
- i 4 \hb ( p_0 - \bar{p}_0 )(x_0 \bar{\si}_0^2 + \bar{x}_0 \si_0^2) }{ 4\hb^2 (\si_0^2 + \bar{\si}_0^2) }
\label{eq: b0}
\\
b_1(t) &=& \frac{ x_0 \bar{\si}_0^2 + \bar{x}_0 \si_0^2 }{ \si_0^2 + \bar{\si}_0^2 } 
+ \frac{ \bar{p}_0 \bar{\si}_0^2 + p_0 \si_0^2 }{ m(\si_0^2 + \bar{\si}_0^2) } \uptau(t)
- i \left[  \frac{ \hb \uptau(t) }{2m} \frac{ x_0 - \bar{x}_0 }{ \si_0^2 + \bar{\si}_0^2 } + 
2 \frac{ (\bar{p}_0 - p_0) \si_0^2 \bar{\si}_0^2 }{ \hb(\si_0^2 + \bar{\si}_0^2) } \right]
 \label{eq: b1}  \\
b_2(t) &=& \frac{ \si_0^2 \bar{\si}_0^2 }{ \si_0^2 + \bar{\si}_0^2 } + \frac{\hb^2 \uptau(t)^2}{4m^2(\si_0^2 + \bar{\si}_0^2)} + \frac{ 4\ga t + 4 e^{-2\ga t} - 3 - e^{-4\ga t} }{ 16m^2\ga^3 }D - i \frac{ \hb (\si_0^2 - \bar{\si}_0^2) }{ 2m(\si_0^2 + \bar{\si}_0^2) } \uptau(t). \label{eq: b2}
\end{eqnarray}
Note that for $ \bar{\si}_0 = \si_0 $ one has $ b_2(t) = \si_t^2 / 2 $ where $\si_t = w_t|_{\eta=0}$ is obtained from Eq. (\ref{eq: wt}) by imposing $\eta=0$. $ P_{11}(x, t) $ and $ P_{22}(x, t) $ are 
given by (\ref{eq: probden}) by using appropriate momenta. 
For $ \bar{\si}_0 = \si_0 $ and $ \bar{x}_0 = x_0 $, corresponding to the {\it one-slit} diffraction, from (\ref{eq: Gamma12}) one obtains
\begin{eqnarray} \label{eq: Gamma12_Gauss}
\Gamma_{12}(t) &=& - \frac{ \si_0^2 (p_0-\bar{p}_0)^2}{2\hb^2} \left\{ 1 - 
\frac{ \si_0^2 }{ \si_t^2 } \left[ 1 + \frac{ \hb^2 }{ 4 m^2 \si_0^4 } \uptau(t)^2 \right] \right\}
\end{eqnarray}
which is certainly negative as can be seen from (\ref{eq: wt}). 
The decoherence function is the same for bosons and fermions.
The decoherence due to the last term of equation (\ref{eq: P_sp}) is here interpreted as loss of being indistinguishable as described in \cite{MoMi-EPJP-2020-1}.
Notice that the case $ p_0 = \bar{p}_0 $ can occur only for bosons which then the wavefunction (\ref{eq: Psi_pm}) takes the product form just as classical states, revealing that quantum statistics is unimportant when the decoherence function $ \Gamma_{12}(t) $ becomes zero.
However, this term vanishes if the overlapping integral is negligible. In such a case the quantum statistics is unimportant. This situation can also happen in isolated systems and it is not a result of interaction with the environment. Therefore, one should consider the effect of environment on $ P_{12}(x, t) $ and $ P_{21}(x, t) $ as an additional source of decoherence taking place for identical particle systems. 

%
%

%
%
%
%

\section{Quantum shutter problem: Diffraction in time}

Dynamics of a beam of particles described initially by a plane wave, confined to the negative semi-infinite region $ x < 0 $, has been studied by Moshinsky \cite{Mo-PR-1952} in the framework of the Schr\"{o}dinger equation when the shutter is suddenly removed at $t=0$. This problem is of fundamental importance in the framework of isolated systems where the so-called  diffraction in time was first reported \cite{CaCaMu-PR-2009}. 

Our aim here is to carry out the same dynamics but in the  CL framework. In order  to simplify this analysis, we restrict ourselves to the case of negligible dissipation which corresponds to time-scales much shorter than the relaxation time $\ga^{-1}$. In this limit, the second term in Eq. (\ref{eq: CL eq}) can be safely neglected. Propagator for the resulting master equation in the free case is expressed as \cite{Ye-PRA-2010}
\begin{eqnarray} \label{eq: prop_free}
	G(x, y, t|x', y', 0) &=& \frac{ m }{ 2\pi \hb t }
	\exp \left[  \frac{i m}{2\hb t} \{ (x-x')^2 - (y-y')^2 \} \right. \nonumber \\
	&-& \left. \frac{D t}{3\hb^2} \{ (x-y)^2 + (x-y)(x'-y') + (x'-y')^2 \} \right]  .
\end{eqnarray}
For $D=0$, this propagator reduces to that of the free particle in the framework of the von Neumann equation. If the initial state is $e^{ikx} \theta(-x)$ and using the center of mass and relative coordinates $R'$ and $r'$ given respectively by (\ref{eq: cm}) and (\ref{eq: rel}) but with $x'$ and $y'$ instead of $x$ and $y$, the diagonal elements of the density matrix are written as
\begin{eqnarray} 
	P(x, t)  \equiv  \rho(x, x, t) & = & \frac{ m }{ 2\pi \hb t } \int_{-\infty}^{0} dR' \int_{2R'}^{-2R'} dr' 
	\exp \left[ - \frac{D t}{3\hb^2} r'^2 + i \left( k - \frac{m}{\hb t} (x-R')  \right) r'  \right] \nonumber  \\
	&=& \frac{ m }{ 2\pi \hb t } \int_{-\infty}^{0} dR' f(x, t, R') \label{eq: Pt_DIT}
\end{eqnarray}
with
\begin{eqnarray} \label{eq: f}
	f(x, t, R') &=& \frac{\hb}{2} \sqrt{ \frac{3\pi}{D t} }
	\left( 
	\erf \left[ \frac{ -4DR t^2 + 3i\hb(\hb kt + m(R-x)) }{ 2 \hb \sqrt{3Dt^3} } \right] - \text{c.c.} \right) \nonumber \\
	&\times& \exp \left[ - \frac{ 3( \hb k t - m (x-R) )^2 }{ 4 D t^3 } \right] 
\end{eqnarray}
where erf$(\cdot)$ means the error function and c.c inside the parenthesis the complex conjugate of the first term. Integral (\ref{eq: Pt_DIT}) can not be carried out analytically and should be obtained numerically.
In the limit $ T \rightarrow 0 $, the above equations read
\begin{eqnarray} 
	f(x, t, R') &=& - \frac{2\hb t}{\hb k t + m (R'-x)} \sin \left[ \frac{2R'(\hb k t + m (R'-x))}{\hb t} \right]
	\\
	P(x, t) &=& \frac{1}{2} \left( C(\xi) + \frac{1}{2} \right)^2 + \frac{1}{2} \left( S(\xi) + \frac{1}{2} \right)^2, \qquad \xi = \sqrt{\frac{m}{\pi\hb t}} \left( \frac{\hb k}{m} t - x \right)
	\label{voN_shutter}
\end{eqnarray}
$ C(\xi) $ and $ S(\xi) $ being the Fresnel integrals. Eq.(\ref{voN_shutter}) is a known result in the context of closed  quantum systems. The well known time oscillations in the density profile are gradually killed by the interaction with the environment.

\section{Results and discussion}

\begin{figure} 
	\centering
	\includegraphics[width=12cm,angle=-0]{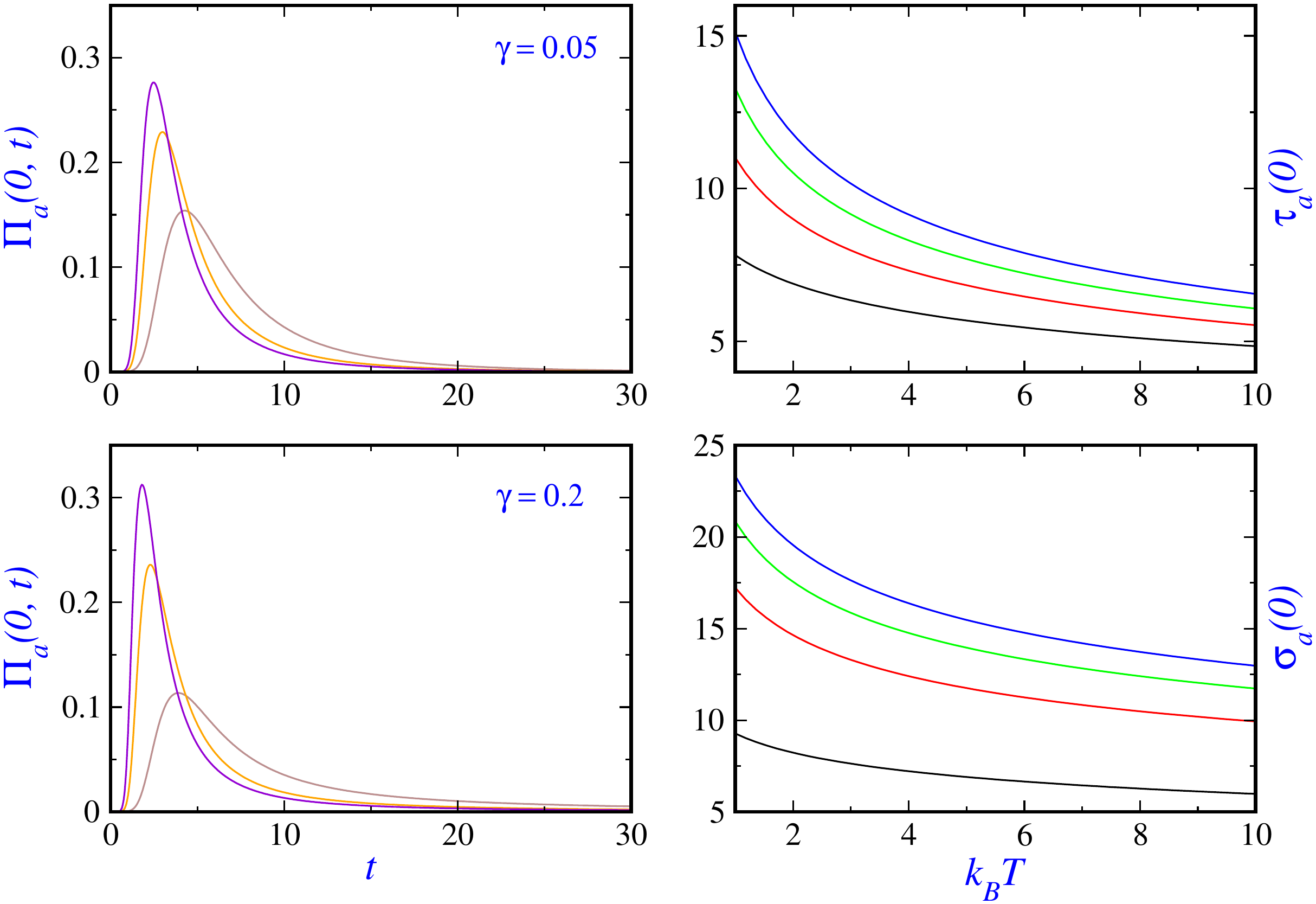}
	\caption{
		Arrival time distributions $ \Pi_a(X, t) $ at the detector location $ X=0 $ for $\ga = 0.05$ (left top panel) and $\ga = 0.2$ (left bottom panel) for different temperatures: $ k_B T = 1 $ (brown curve), $ k_B T = 5 $ (orange curve) and $ k_B T = 10 $ (violet curve). In the right panels, mean arrival times (top) and rms width of the arrival time distribution (bottom) at the detector location have been plotted versus temperature for different values of damping constant: $\ga = 0.05$ (black curve), $\ga = 0.1$ (red curve), $\ga = 0.15$ (green curve) and $\ga = 0.2$ (blue curve). The initial state is chosen to be a minimum-uncertainty-product Gaussian wave packet with parameters $\si_0=1$, $x_0=-5$, $p_0=0.5$, in the absence of external force.
}
	\label{fig: artime} 
\end{figure}

Numerical calculations along this work are given in units of $\hb=1$ and $m=1$. The first aspect we want to analyze is time arrivals from the solutions of Eqs. (\ref{eq: CL eq}) and (\ref{eq: con_CL}). For this goal, we present
arrival time distributions at the origin from Eq. (\ref{eq: ardis})  for propagation of a minimum-uncertainty-product Gaussian wave packet coming from the left . In figure \ref{fig: artime}, these distributions are plotted  for $\ga = 0.05$ (left top panel) and $\ga = 0.2$ (left bottom panel) for different temperatures, $ k_B T = 1 $ (brown curve), $ k_B T = 5 $ (orange curve) and $ k_B T = 10 $ (violet curve). 
In the right panels, mean arrival times, Eq. (\ref{eq: meanar}), and rms width of the distribution, (\ref{eq: sigma_ar}), are displayed versus temperature for different relaxation rates,
$\ga = 0.05$ (black curve), $\ga = 0.1$ (red curve), $\ga = 0.15$ (green curve) and $\ga = 0.2$ (blue curve).
The peaks of these arrival time distributions move to shorter times as temperature increases. As expected, the mean arrival time plotted in the right top panel increases with the relaxation rate for a given temperature. This mean time is also a decreasing function of the temperature, independently on the value of the relaxation rate.
The same behavior is observed for the width of these distributions in the right bottom panel.  
\begin{figure} 
	\centering
	\includegraphics[width=12cm,angle=-0]{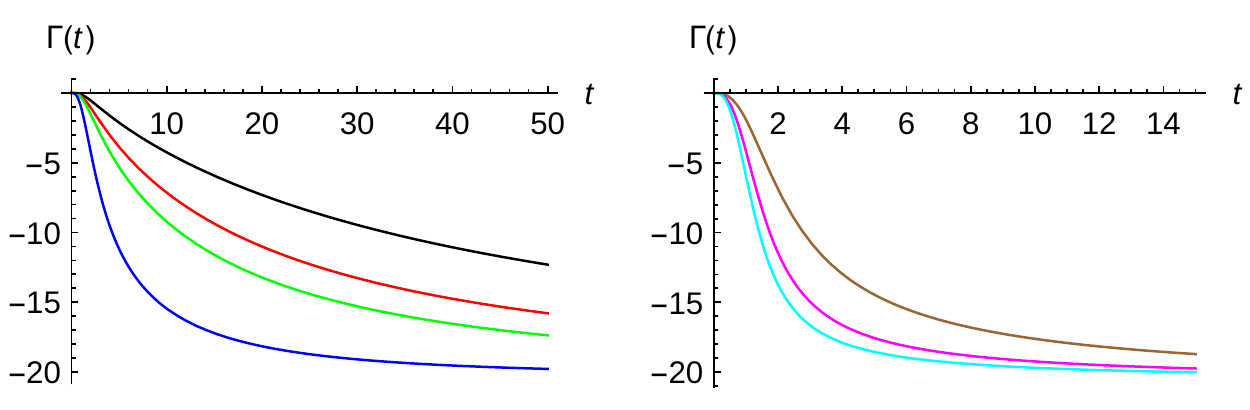}
	\caption{Decoherence function $ \Gamma(t) $, given by Eq. (\ref{eq: deco_func}), versus time for $ k_B T = 1 $ (left panel) and for $ \ga = 0.05 $ (right panel). Curve color codes in the left panel are: $ \ga = 0.005 $ (black), $ \ga = 0.01 $ (red), $ \ga = 0.015 $ (green), $ \ga = 0.05 $ (blue); whereas in the right panel, $ k_B T = 2 $ (brown), $ k_B T = 5 $ (magenta) and $ k_B T = 8 $ (cyan). Parameters for the two minimum-uncertainty Gaussian wave packets are $\si_0=1$, $x_0=5$ and $p_0=-2$. }
	\label{fig: decohere_func} 
\end{figure}

The second aspect to be analyzed is the role of the friction or relaxation rate and temperature for two distinct Gaussian wave packet dynamics, minimum-uncertainty and non-minimum-uncertainty product Gaussian wave packets.
The decoherence function in the interference of two minimum-uncertainty Gaussian  wave packets is studied here for a zero constant field, $g=0$, with parameters $p_0=-2$, $\si_0=1$ and $x_0=5$ according to Eq.(\ref{eq: wf0_sup}). 
With these values for the parameters of the initial wave packets one ensures  negligible initial overlap between them. 
\begin{figure} 
	\centering
	\includegraphics[width=12cm,angle=-0]{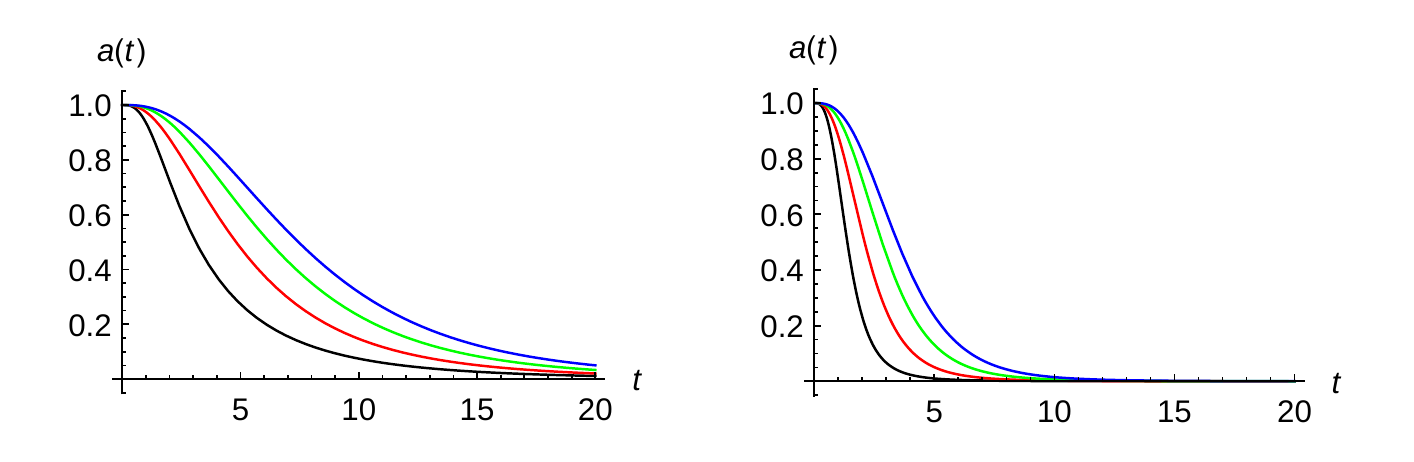}
	\caption{Coherence attenuation coefficient $ a(t) $, given by Eq. (\ref{eq: atten}), versus time for $ \ga=0.005 $; and for $ k_B T = 1 $ (left panel) and $ k_B T = 5 $ (right panel) for different values of stretching parameter; $ \eta = 0 $ (black curve), $ \eta = 1 $ (red curve), $ \eta = 2 $ (green curve) and $ \eta = 3 $ (blues curve). Same parameters as in Figure \ref{fig: decohere_func}. }
	\label{fig: attenuation_func} 
\end{figure}
Figure \ref{fig: decohere_func} displays the decoherence function versus time for a given value of temperature $ k_B T = 1 $ but different relaxation rates (left panel) and for a given value of the damping constant $ \ga = 0.05 $ but different temperatures (right panel). 
Curve color codes in the left panel are as follows: $ \ga = 0.005 $ (black), $ \ga = 0.01 $ (red), $ \ga = 0.015 $ (green), $ \ga = 0.05 $ (blue); and 
in the right panel: $ k_B T = 2 $ (brown), $ k_B T = 5 $ (magenta) and $ k_B T = 8 $ (cyan). The decoherence function is negative and decreases 
with time for both $\gamma$ and $T$, reaching a stationary value in all cases. Thus, the coherence behavior as well as the interference pattern 
is lost gradually with time; in other words, the decoherence process is established gradually with time reaching a stationary value. However, with $T$, the decoherece function decreases faster than with $\ga$, total decoherence being  
reached at much shorter times. Figure \ref{fig: attenuation_func} depicts the coherence attenuation coefficient (\ref{eq: atten}) versus time for the cat state built from stretched wave packets for  $\ga=0.005$ and $k_B T = 1$ (left panel) and $ k_B T = 5 $ (right panel) for different values of the stretching parameter:
$ \eta = 0 $ (black curve), $ \eta = 1 $ (red curve), $ \eta = 2 $ (green curve) and $ \eta = 3 $ (blues curve). As one clearly sees, positive values of $\eta$ reduces the rate of decoherence and this reduction is more pounced for lower temperatures.  
\begin{figure}
	\centering
	\includegraphics[width=15cm,angle=-0]{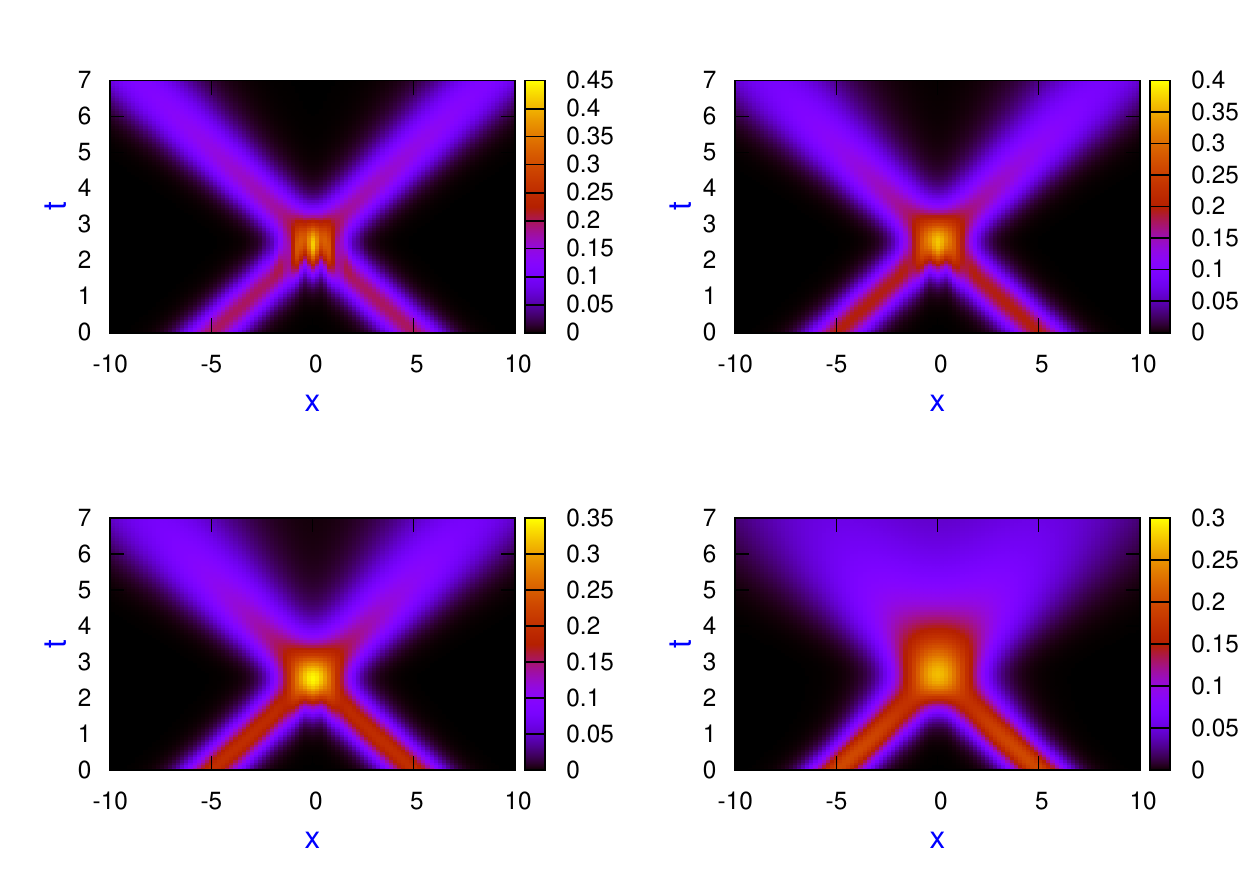}
	\caption{
		Probability density plots (\ref{eq: probden_sup2}) for the superposition of two minimum-uncertainty-Gaussian wave packets, $ \eta = 0 $, for $ k_B T = 1 $ and different values of the relaxation rate: 
		$\ga=0.0001$ (left top panel), $\ga=0.001$ (right top panel), $\ga=0.003$ (left bottom panel) and $\ga=0.01$ (right bottom panel).  
		Same parameters as in Figure \ref{fig: decohere_func}. 
	}
	\label{fig: den_sup_Gauss} 
\end{figure}

\begin{figure} 
	\centering
	\includegraphics[width=10cm,angle=-0]{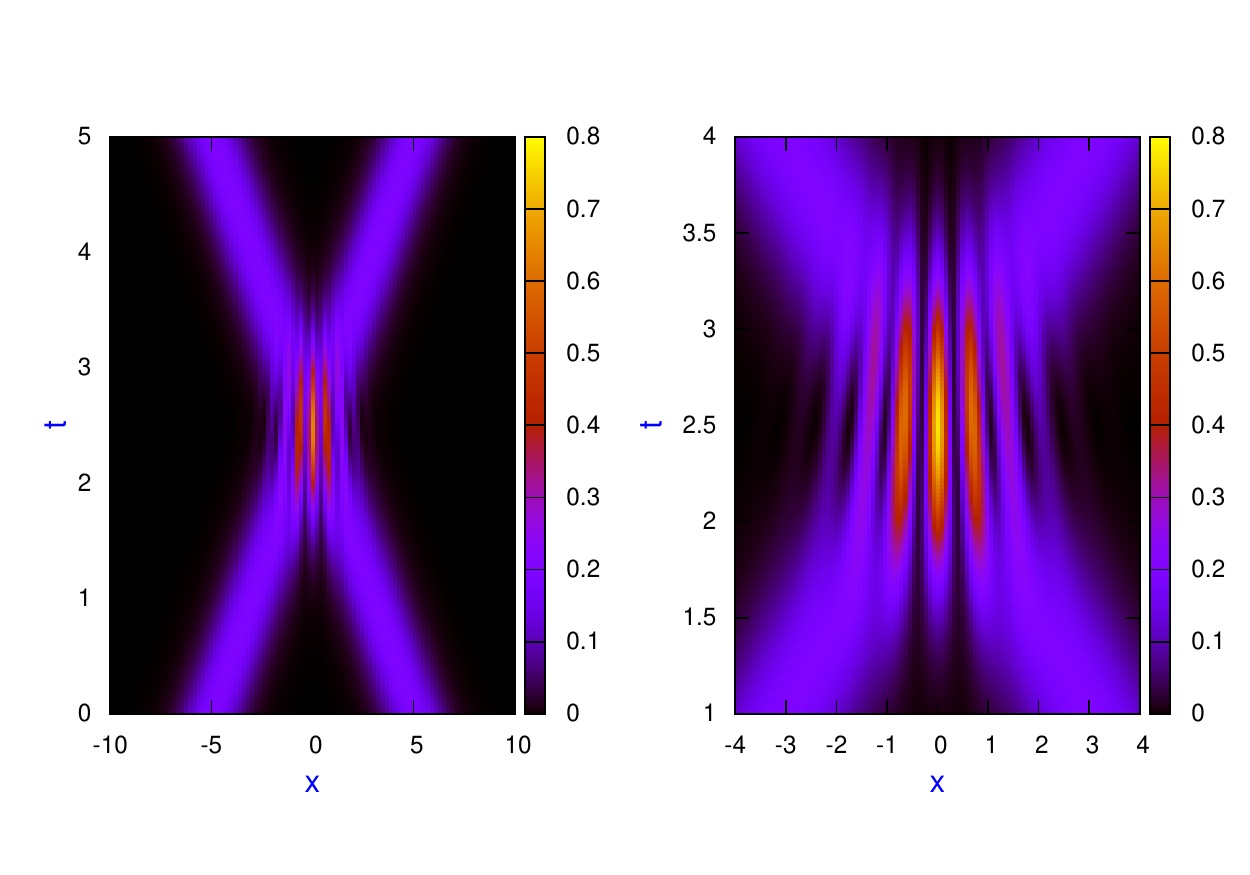}
	\caption{
		Probability density (\ref{eq: probden_sup2}) plots for $ k_B T = 0 $. This corresponds to the evolution of the initial state (\ref{eq: wf0_sup}) issued from the Schr\"{o}diger equation. The right panel is a zoom-out of the left one around the interference region. Same parameters as in Figure \ref{fig: decohere_func}.
	}
	\label{fig: den_sup_Gauss_Sch} 
\end{figure}

\begin{figure}
	\centering
	\includegraphics[width=12cm,angle=-0]{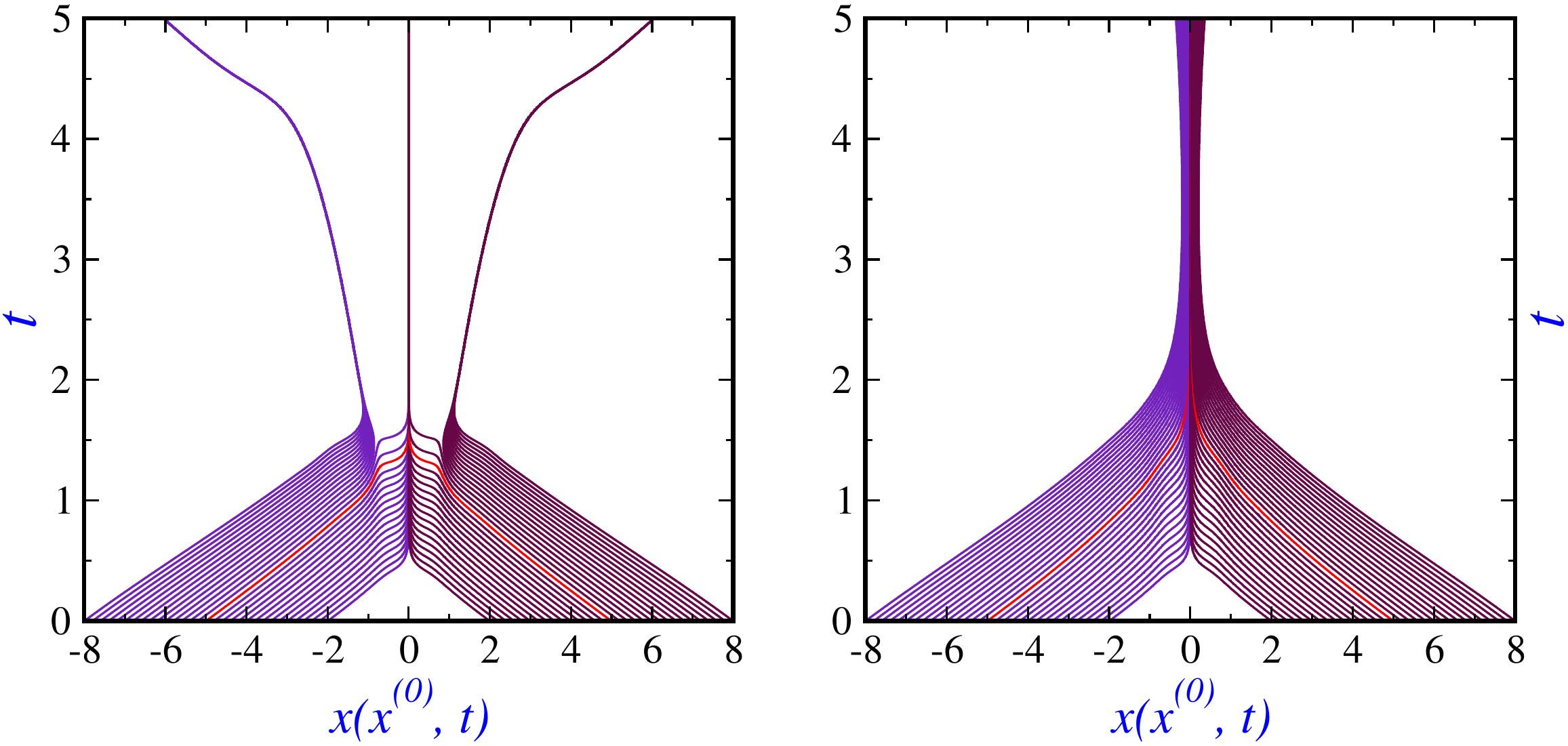}
	\caption{
		A selection of Bohmian trajectories for the superposition of the two minimum-uncertainty-Gaussian wave packets for $ k_B T = 1 $ and for two different values of the relaxation coefficient:  $\ga=0.001$ (left panel) and  $\ga=0.05$ (right panel). Indigo (maroon) trajectories correspond to Bohmian trajectories running at the left (right) wave packet. Red trajectories correspond to Bohmian ones following the center of the wavepackets. Same parameters as in Figure \ref{fig: decohere_func}.
	}
	\label{fig: trajs} 
\end{figure}

To better illustrate the effect of damping on the interference pattern, in Fig. \ref{fig: den_sup_Gauss}  probability density plots  (\ref{eq: probden_sup2}) are shown for $ k_B T = 1 $ and different  
relaxation rates: $\ga=0.0001$ (left top panel), $\ga=0.001$ (right top panel), $\ga=0.003$ (left bottom panel) and  $\ga=0.01$ (right bottom panel). The interference pattern is drastically reduced as seen in the right bottom panel.
For comparison, we have also plotted  the probability density for the corresponding isolated system described by the Schr\"{o}dinger equation in Fig. \ref{fig: den_sup_Gauss_Sch} for $ k_B T = 0 $ and $\ga=0$. The right panel is a zoom-out of the left one around the interference region. Parameters of the initial wave packets have been chosen in such a way that they do not spread substantially during the time evolution. 
Bohmian trajectories have bee plotted in Fig. \ref{fig: trajs} for $ k_B T = 1 $ and two different values of $\ga$
: $\ga=0.001$ (left panel) and  $\ga=0.05$ (right panel). 
Indigo (maroon) trajectories correspond to Bohmian trajectories running for the left (right) wave packet. Red trajectories correspond to Bohmian ones following the center of the wave packets. Same parameters as in Figure \ref{fig: decohere_func}.
As this figure shows for the relaxation coefficient $\ga=0.001$, these trajectories converge to three distinct bunches corresponding to the interference fringes in Fig. \ref{fig: den_sup_Gauss}. However, at higher $\ga$ values, the fringes have disappeared. 

\begin{figure}
\centering
\includegraphics[width=15cm,angle=-0]{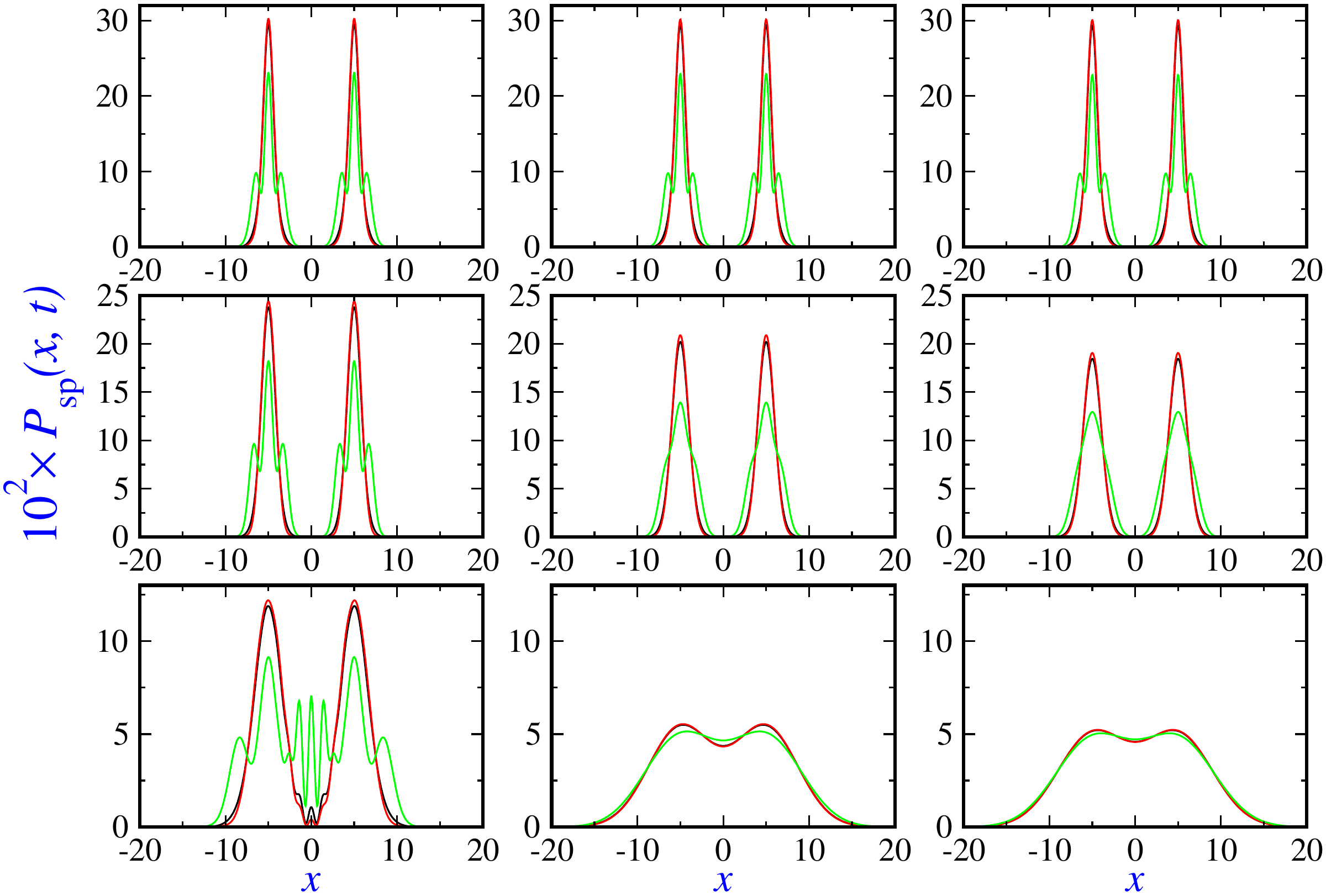}
\caption{
Scaled single-particle probability density versus $x$ (\ref{eq: P_sp}) for two identical spin-less bosons (red) and fermions (green) for an ideal closed system (left column). The same is plotted for  $ k_B T = 10 $ but different damping constants $\ga = 0.2 $ (middle column) and $\ga = 0.4 $ (right column) at different times: $ t = 0.1 $ (top row), $ t = 0.5 $ (middle row) and $ t = 2 $ (bottom row).
For comparison, distinguishable particles obeying classical Maxwell-Boltzmann statistics, this probability is also shown by black curves. 
Both one-particles states $\psi$ and $\phi$ have been taken as superposition of two motionless minimum-uncertainty-product-Gaussian wavepackets, (\ref{eq: wf0_sup}), with $x_0 = \pm 5$ but different widths $\si_0 = 1$ and $\delta_0 = 0.5$. 
}
\label{fig: rhosp1} 
\end{figure}

The third aspect  deals with the interference of identical particles when an open dynamics is considered; in particular, the role played by the temperature. 
Figure \ref{fig: rhosp1} shows decoherence in the context of identical particles through the single-particle density (\ref{eq: P_sp}). Here, the effect of friction is considered for a given temperature. Scaled single-particle probability density has been plotted versus distance for two identical spin-less bosons (red) and fermions (green) for a closed system (left column). The same probability is shown for $ k_B T = 10 $ but different damping constants $\ga = 0.2 $ (middle column) and $\ga = 0.4 $ (right column) at different times: $ t = 0.1 $ (top row), $ t = 0.5 $ (middle row) and $ t = 2 $ (bottom row). For comparison, the behavior of distinguishable particles obeying classical Maxwell-Boltzmann statistics is also shown by black curves. This black curve almost coincides with the red one (bosons) for our choice of parameters.
Both one-particles states $\psi$ and $\phi$ have been taken as superposition of two motionless minimum-uncertainty-product-Gaussian wavepackets, (\ref{eq: wf0_sup}), with $x_0 = \pm 5$ but different widths $\si_0 = 1$ and $\delta_0 = 0.5$. 
With $\gamma$, the widths increase with time, for the considered region of time, and quantum statistics becomes less and less important. 
The oscillations observed for fermions in the closed system scenario are totally suppressed by $\gamma$ and $T$.
The space distributions of fermions are always slightly wider than for bosons due to the well-known anti-bunching property displayed by the former. Bimodality of the corresponding single-particle probability densities also tends to disappear with friction and time.
In other words, the symmetry of the wave function is not robust enough to keep the corresponding statistics along time. This feature was also observed previously when only friction was considered \cite{MoMi-EPJP-2020-1} in a different context, the Caldirola-Kanai approach.

\begin{figure}
	\centering
	\includegraphics[width=15cm,angle=-0]{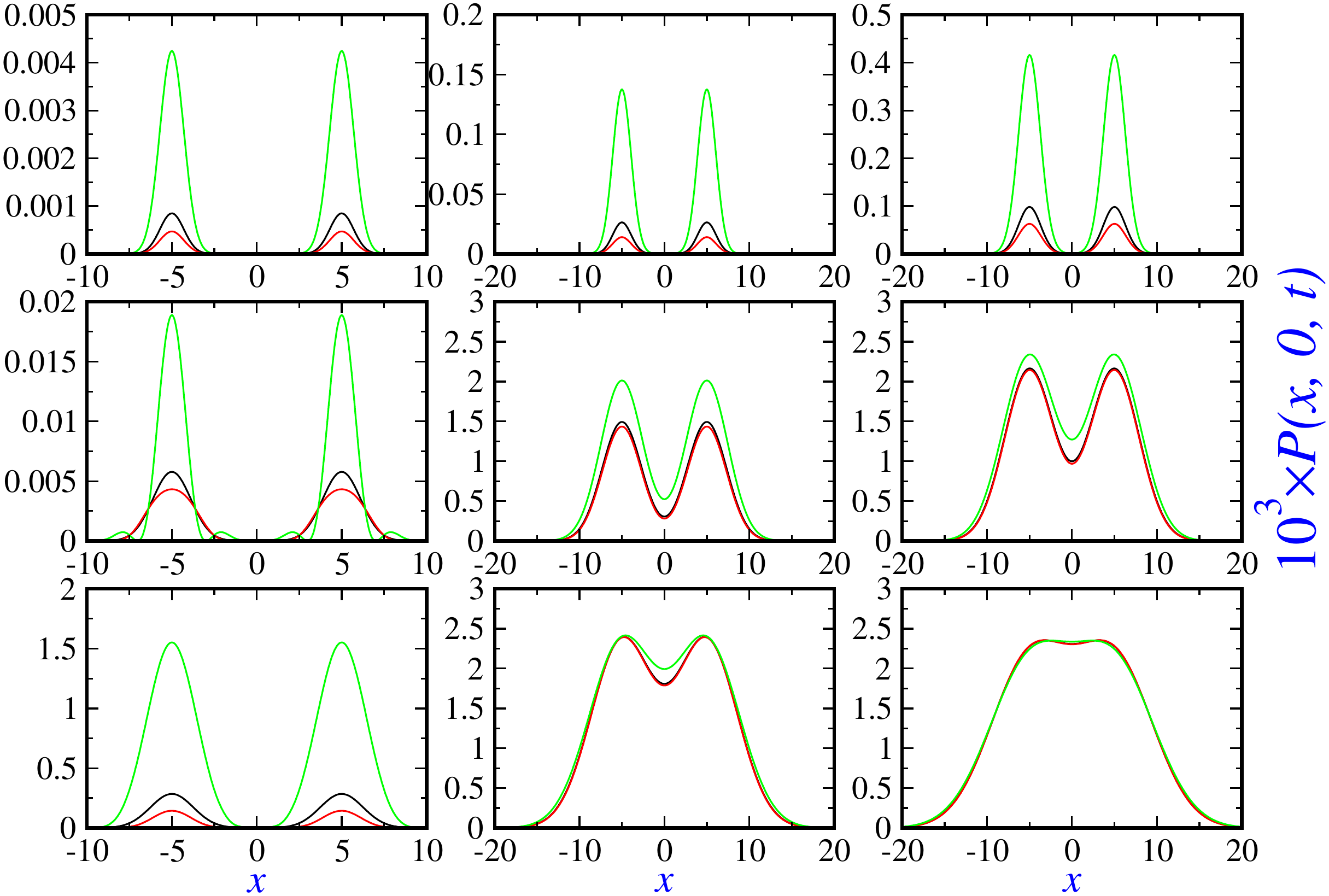}
	\caption{
		Scaled joint detection probability versus $x$ (\ref{eq: P_2p}) for two spin-less distinguishable particles (black) identical bosons (red) and fermions (green) for a closed system (left column). One of the two particles is at the origin. This probability is also shown for $ \ga=0.4 $ but different temperatures $k_B T = 15$ (middle column) and $k_B T = 25$ (right column) at different times; $ t = 0.5 $ (top row), $ t = 1 $ (middle row) and $ t = 1.5 $ (bottom row).
		Both one-particles states $\psi$ and $\phi$ has been taken as superposition of two motionless minimum-uncertainty-product-Gaussian wavepackets, (\ref{eq: wf0_sup}), with $x_0 = \pm 5$ but different widths $\si_0 = 1$ and $\delta_0 = 0.5$. 
	}
	\label{fig: rhosp2} 
\end{figure}

In figure \ref{fig: rhosp2}, the joint detection probability for finding both particles with one at the origin, has been depicted for a given damping constant but different temperatures; scaled joint detection probability versus distance (\ref{eq: P_2p}) for two spin-less distinguishable particles (black) identical bosons (red) and fermions (green) for a closed system (left column). This probability is also plotted for $ \ga=0.4 $ but different temperatures $k_B T = 15$ (middle column) and $k_B T = 25$ (right column) and at different times, $ t = 0.5 $ (top row), $ t = 1 $ (middle row) and $ t = 1.5 $ (bottom row).
Both one-particles states $\psi$ and $\phi$ have been taken again as superposition of two motionless minimum-uncertainty-product-Gaussian wavepackets, (\ref{eq: wf0_sup}), with $x_0 = \pm 5$ but different widths $\si_0 = 1$ and $\delta_0 = 0.5$. For the closed dynamics, the corresponding probability of finding the second particle at the origin is nearly zero.
However, with $T$ and time this probability is small but different from zero. Furthermore, the  widths also increase with $T$ and time; and the joint detection distribution are wider and wider, being wider for fermions than for bosons. The same features are then observed in this analysis. 

%

%
\begin{figure} 
	\centering
	\includegraphics[width=12cm,angle=-0]{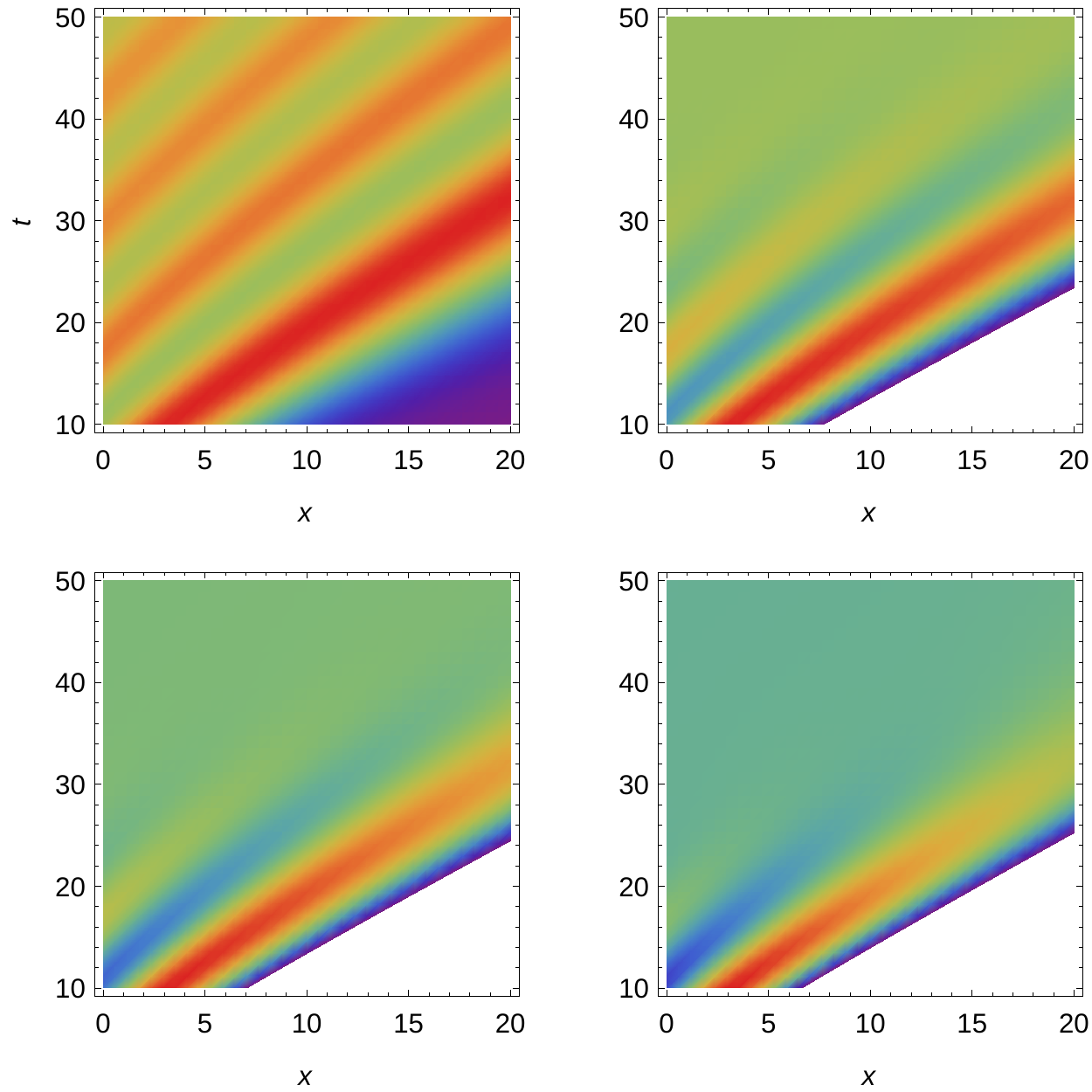}
	\caption{
		Density plots of the probability density $ P(x, t) $ for the released beam $ e^{i k x} $ in the shutter problem for isolated closed system (left top panel) and the open system in the context of the CL framework for $ \ga = 0.0001 $ and for different values of temperature: $k_B T = 1 $ (right top panel), $k_B T = 2 $ (left bottom panel) and $k_B T = 4 $ (right bottom panel). Momentum is $k=1$. 
	}
	\label{fig: DIT_denplot} 
\end{figure}
\begin{figure} 
	\centering
	\includegraphics[width=12cm,angle=-0]{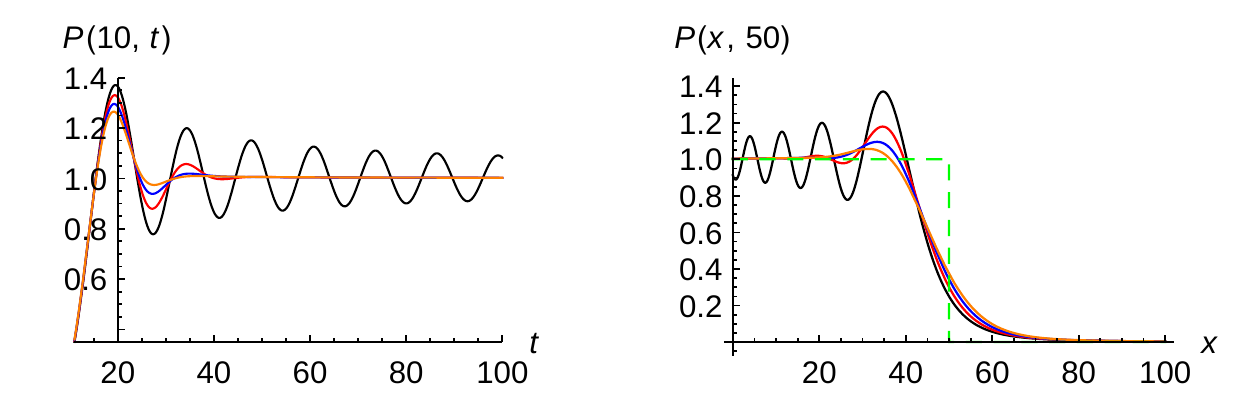}
	\caption{
		Probability density $ P(x, t) $ for the released beam $ e^{i k x} $ in the shutter problem, in the negligible dissipation limit, versus time at the space coordinate $ x=10 $ (left panel) and versus $x$ at $t=50$ (right panel) for $ k_B T = 2 $; and $ \ga = 0.00005 $ (red curves), $ \ga = 0.0001 $ (blue curves) and $ \ga = 0.00015 $ (orange curves). Black curves refer to the the same problem but in the framework of isolated systems described by the von Neumann equation and the dashed green shows the classical pattern. Momentum is $k=1$. 
	}
	\label{fig: DIT_shutter} 
\end{figure}

Finally, the last aspect studied here is the problem of sudden release of the initially confined plane wave from a hard wall which is known as the quantum shutter problem. For fulfilling the negligible dissipation limit we take $ \ga = 0.0001 $  in the time region $ t \leq 100 $. 
For a given $\ga$ and $T$, as time increases, oscillations of the function $f_2(x, t, R')$ in Eq. (\ref{eq: f}) are drastically reduced; in particular, when $ R \rightarrow -\infty $. For our parameters replacing the lower limit of the integral by $-200$ suffices and yields reasonable results. 
Figure \ref{fig: DIT_denplot} depicts probability densities (\ref{voN_shutter}) for the isolated system (left top panel) and (\ref{eq: Pt_DIT}) for the open system for $ \ga = 0.0001 $ and different temperatures (the remaining panels) in the space and time regions $ 0 \leq x \leq 20 $ and $ 10 \leq t \leq 50 $, respectively. 
While diffraction in both space and time is seen in the isolated case, oscillations are gradually killed in the presence of environment as temperature increases (red curves). In figure \ref{fig: DIT_shutter}, the probability densities versus time at a given place (left panel) and versus space coordinate at a given time (right panel) have been plotted to show explicitly oscillations around the classical pattern.   
Black curves refer to the the same problem but in the framework of isolated systems described by the von Neumann equation, the dashed green shows the classical pattern. As one sees oscillations are gradually killed in the presence of environment as the relaxation rate increases for a given temperature.

\section{ Concluding remarks }

In this work, we have explored several new aspects of decoherence within the CL framework; in particular, arrival time distributions,  identical spinless particles and diffraction in time. The role of a positive stretching parameter in the Gaussian wave packets has also been investigated leading to a reduction of the decoherence rate. As mentioned previously, the source of decoherence  is in the last term of  Eq. (\ref{eq: CL eq}) where the diffusion coefficient depends on the temperature and friction or damping rate. Here the decoherence is analyzed in terms of the so-called decoherence function and attenuation coefficient when typical interference pattern expressions are studied for indistinguishable and distinguishable particles.

For identical particles, we have clearly shown that the symmetry of the wave function is not robust enough to keep the statistical signature of the particles leading to a gradual decoherence. This can be seen as a gradual lost of being indistinguishable when the temperature and friction are present. The corresponding analysis for arrival time distributions as well the diffusion process should also be investigated. 

Finally, the decoherence process has also been observed in time itself, in the well-known process of diffraction in time. A gradual reduction of the oscillations which are the hallmark of this type of diffraction is clearly seen with temperature and low frictions.


\appendix

\section{Continuity equation for the single-particle density} \label{app: con-eq}

Taking time derivative of both sides of Eq. (\ref{eq: P_sp}) yields
\begin{eqnarray} 
	\frac{\pa}{\pa t} P_{\sip, \pm}(x, t) &=& \mathcal{N}_{\pm}^2 \left\{  
	\frac{\pa}{\pa t}P_{11}(x, t) + \frac{\pa}{\pa t}P_{22}(x, t) 
	\pm \re \{
	\int dx' \frac{\pa}{\pa t} [  P_{12}(x, t) P_{21}(x', t) ] \} \right\} 
\end{eqnarray}
Now using the continuity equation (\ref{eq: con_CL}) and imposing appropriate boundary condition yielding zero  boundary terms we obtain the following continuity equation,
\begin{eqnarray} \label{eq: con_sp}
	\frac{\pa}{\pa t} P_{\sip, \pm}(x, t) + \frac{\pa}{\pa x} J_{\sip, \pm}(x, t) &=& 0 
\end{eqnarray}
where 
\begin{eqnarray}
	J_{\sip, \pm}(x, t) &=& \mathcal{N}_{\pm}^2 [ J_{11}(x, t) + J_{22}(x, t) \pm \re\{J_{12}(x, t) s(t)\} ]
\end{eqnarray}
and $ J_{kl}(x, t) $ given by diagonal elements of (\ref{eq: cur_den_mat}) satisfies Eq. (\ref{eq: con_CL}) with $ P_{kl}(x, t) $.


\vspace{1cm}
\noindent
{\bf Acknowledgements}
SVM acknowledges support from the University of Qom and SMA  from  Fundaci\'on Humanismo y Ciencia.



\end{document}